\documentclass[sigconf,authorversion,nonacm]{acmart}

\usepackage{balance}

\usepackage[utf8]{inputenc}
    \usepackage{listings,xcolor}
    \usepackage[T1]{fontenc}
    \usepackage{xcolor}
    \usepackage[scaled=0.7]{DejaVuSansMono}
\definecolor{commentgreen}{RGB}{2,112,10}
\definecolor{eminence}{RGB}{108,48,130}
\definecolor{weborange}{RGB}{255,165,0}
\definecolor{frenchplum}{RGB}{129,20,83}

\lstdefinelanguage{elixir}{
    morekeywords={case,catch,def,do,else,false,%
        use,alias,receive,timeout,defmacro,defp,%
        for,if,import,defmodule,defstruct,defprotocol,%
        nil,defmacrop,defoverridable,defimpl,%
        super,fn,raise,true,try,end,with,%
        unless},
    otherkeywords={<-,->, |>, \%\{, \}, \{, \, (, )},
    sensitive=true,
    morecomment=[l]{\#},
    morecomment=[n]{/*}{*/},
    morecomment=[s][\color{purple}]{:}{\ },
    morestring=[s][\color{blue}]"",
    commentstyle=\color{commentgreen},
    keywordstyle=\color{eminence},
    stringstyle=\color{red},
	basicstyle=\ttfamily\linespread{0.8}\selectfont,
	breaklines,
	showstringspaces=false,
	frame=tb
}
	
\AtBeginDocument{%
  \providecommand\BibTeX{{%
    \normalfont B\kern-0.5em{\scshape i\kern-0.25em b}\kern-0.8em\TeX}}}

\setcopyright{acmcopyright}
\copyrightyear{2022}
\acmYear{2022}
\acmDOI{XXXXXXX.XXXXXXX}

\acmConference[ICPC '22 ERA]{30th IEEE/ACM International Conference on Program Comprehension}{May 16--17,
2022}{Pittsburgh, PA}
\acmPrice{15.00}
\acmISBN{978-1-4503-XXXX-X/18/06}

 \usepackage{array,multirow,graphicx}
 \usepackage{float}

\begin{document}

\title{Code Smells in Elixir: Early Results from a Grey Literature Review}



\author{Lucas Francisco da Matta Vegi}
\email{lucasvegi@dcc.ufmg.br}
\author{Marco Tulio Valente}
\email{mtov@dcc.ufmg.br}
\affiliation{%
  \institution{Federal University of Minas Gerais (UFMG)}
  \city{Belo Horizonte}
  \state{Minas Gerais}
  \country{Brazil}
}



\renewcommand{\shortauthors}{Vegi and Valente.}

\begin{abstract}
  Elixir is a new functional programming language whose popularity is rising in the industry. However, there are few works in the literature focused on studying the internal quality of systems implemented in this language. Particularly, to the best of our knowledge, there is currently no catalog of code smells for Elixir. Therefore, in this paper, through a grey literature review, we investigate whether Elixir developers discuss code smells. Our preliminary results indicate that 11 of the 22 traditional code smells cataloged by Fowler and Beck are discussed by Elixir developers. We also propose a list of 18 new smells specific for Elixir systems and investigate whether these smells are currently identified by Credo, a well-known static code analysis tool for Elixir. We conclude that only two traditional code smells and one Elixir-specific code smell are automatically detected by this tool. Thus, these early results represent an opportunity for extending tools such as Credo to detect code smells and then contribute to improving the internal quality of Elixir systems.
\end{abstract}

\ccsdesc[500]{Software and its engineering~Designing software}
\ccsdesc[300]{Software and its engineering~Software development techniques}

\keywords{Code Smells, Elixir, Functional Programming}

\maketitle

\section{Introduction}

Elixir is a modern functional programming language that has high performance in parallel and distributed environments. The language has an extensible and friendly syntax, similar to the Ruby language. With Elixir, developers can create scalable code transparently without worrying so much about synchronization problems in multithreaded environments \cite{Thomas18}. Currently, more than 600 companies around the world---including Pinterest, Discord, Adobe, and Spotify---are already using Elixir in their production code.\footnote{\href{https://elixir-companies.com/en}{https://elixir-companies.com/en}}


As with any programming language, it is natural to expect that Elixir developers make bad design decisions and then implement sub-optimal code structures, known as code smells \cite{fowler1999refactoring}. These structures decrease the internal software quality, impairing maintainability \cite{Yamashita13}\cite{Soh16} and increasing bug-proneness \cite{LiAndShatnawi07}\cite{Olbrich10}.

However, to the best of our knowledge, there are no papers in the literature focused on studying the internal quality of Elixir systems or, in more general terms, of systems implemented in functional languages. For example, Sobrinho et al. \cite{Sobrinho2021} conducted a systematic review of articles on code smells published between 1990 and 2017. None of the 104 smells reviewed in their paper refer to functional languages.
We also performed a Google Scholar search and did not find more recent code smells papers (2018-2022) that directly consider functional languages such as Elixir.


In order to fill this gap, in this paper, we investigate whether Elixir developers have discussions on code smells. Since Elixir is a new language, we used in our research a grey literature review. This methodology was chosen because it is a good source of information about emerging technologies \cite{KAMEI2021}\cite{Zhang2020ICSE}. 

Our contributions are threefold: (1) we find that half of the traditional code smells also generate discussions among Elixir developers; (2) we catalog 18 novel Elixir-specific code smells and classify them into two different groups (\textit{design-related smells} and \textit{low-level concerns smells}); and (3) we report that Credo,\footnote{\href{http://credo-ci.org/}{http://credo-ci.org/}} the main linter tool for Elixir, is able to identify only three code smells out of 40 smells, considering traditional and novel ones. This finding motivates the implementation of new code smell detection tools in future work.

 The remainder of this paper is organized as follows. In \hyperref[sec:background]{Section II}, we present background information on Elixir and code smells. In \hyperref[sec:methodology]{Section III}, we present our research questions and methods. We detail and discuss our early results in \hyperref[sec:results]{Section IV}. In \hyperref[sec:threats]{Section V}, we list threats to validity. Finally, we present related work in \hyperref[sec:relatedWork]{Section VI} and ideas for future work in \hyperref[sec:futureWork]{Section VII}.


\section{Background}
\label{sec:background}


\subsection{Elixir}
\label{sec:elixirSubsection}

Elixir programs are composed by {\tt modules}, which are groups of functions. Next, we show a simple Elixir module ({\tt Circle}) composed by two functions---{\tt area} and {\tt circumference}. In lines 10-11, we also show two calls of these functions, using Elixir's interactive shell (IEx).\footnote{\href{https://hexdocs.pm/iex/1.13/IEx.html}{https://hexdocs.pm/iex/1.13/IEx.html}}\\[-0.2cm]

\begin{lstlisting}[language=elixir] 
defmodule Circle do
  def area(radius) do
     3.14 * (radius * radius)
  end
  def circumference(radius) do
     2 * 3.14 * radius
  end
end
...
iex(1)> Circle.area(15)          # 706.5
iex(2)> Circle.circumference(15) # 94.2
\end{lstlisting}

Modules can also define {\tt structs}, which are key-value pairs similar to objects. We show next an Elixir module with a struct that represents a point in a cartesian plane. This struct has two fields---{\tt x} and {\tt y}---that are initialized with {\tt nil} (line 2). In Elixir, a struct has the same name of the module where it is defined . Besides a struct, {\tt Point} has two functions---{\tt distance} and {\tt move}. A pipe operator ({\tt |>}) is used in the {\tt distance} function (line 5) to express function composition. For example, {\tt f() |> g(p)} is equivalent to {\tt g(f(), p)}. Finally, it is important to mention that structs are immutable data structures. For this reason, the {\tt move} function creates a new {\tt Point} with the new location, instead of changing the coordinates of the current {\tt Point} (line 8). Specifically, in Elixir, {\tt \%T\{...\}} is similar to a {\tt new T(...)} in a mainstream object-oriented language.\\[-0.2cm] 

\begin{lstlisting}[language=elixir]
defmodule Point do
  defstruct [x: nil, y: nil]
  def distance(p1, p2) do
     Float.pow(p2.x - p1.x, 2) + Float.pow(p2.y - p1.y, 2)
     |> Float.pow(0.5)
  end
  def move(p, delta_x, delta_y) do
     %Point{x: p.x + delta_x, y: p.y + delta_y}
  end
end
...
iex(1)> p1 = %Point{x: 2.0, y: -3.0} #struct creation
iex(2)> p2 = %Point{x: 4.0, y: 5.0}  #struct creation
iex(3)> Point.distance(p1, p2)  # 8.246211251235321
iex(4)> Point.move(p1, 5, 4)    # %Point{x: 7.0, y: 1.0}
\end{lstlisting}

Elixir programs are executed by BEAM, which is the virtual machine  designed to run Erlang programs.\footnote{\href{https://www.erlang.org/}{https://www.erlang.org/}} BEAM is known to be a fault-tolerant and powerful environment to run distributed systems \cite{Almeida18}. Therefore, Elixir programs can easily interoperate with Erlang systems.

\subsection{Code Smells}
\label{sec:codesmellsSubsection}

Fowler and Beck \cite{fowler1999refactoring} coined the terms {\em code (or bad) smells} to name low quality code structures that can compromise software maintenance and evolution. In the literature, other names are also used with the same purpose, such as anti-patterns \cite{brown1998antipatterns}, code anomalies \cite{OizumiEtAl14codeAnomalies}, and bad practices \cite{TaibiLenarduzzi18}. 


We performed an initial analysis of the 22 code smells proposed by Fowler and Beck. Although they have been proposed in the nineties, for object-oriented languages, such as Java, we concluded that 20 smells (out of 22) also apply to Elixir. Only  \textit{Parallel Inheritance Hierarchies} and \textit{Refused Bequest} are not compatible with Elixir structures and semantics. The reason is that Elixir does not have inheritance and these smells are tightly related to this mechanism. On the other side, smells such as  \textit{Duplicated Code}, \textit{Long Function}, \textit{Feature Envy}, \textit{Shotgun Surgery}, and others can be directly translated to Elixir code. 


However, this initial analysis is not grounded in the practice. Therefore, we cannot affirm that Elixir developers are aware and concerned by the code smells originally proposed by Fowler and Beck. Furthermore, Elixir might have its own code smells, which are specific to the language syntax, semantics, and paradigm.

\section{Methodology}
\label{sec:methodology}

Since Elixir is a new programming language, we have few scientific articles investigating software engineering and quality aspects of Elixir systems. For this reason, the grey literature---composed of blogs, forums, videos, etc---is an interesting source of information for our goals in this study \cite{KAMEI2021}\cite{Zhang2020ICSE}. Therefore, we decided to conduct a grey literature review to answer three key research questions:

\begin{itemize}
    \item \textbf{RQ1:} Do Elixir developers discuss traditional code smells?
    \item \textbf{RQ2:} Do Elixir developers discuss other smells?
    \item \textbf{RQ3:} Does a well-known static code analysis tool for Elixir detect code smells?
\end{itemize}

\begin{figure}[!ht]
  \centering
  \includegraphics[width=\linewidth]{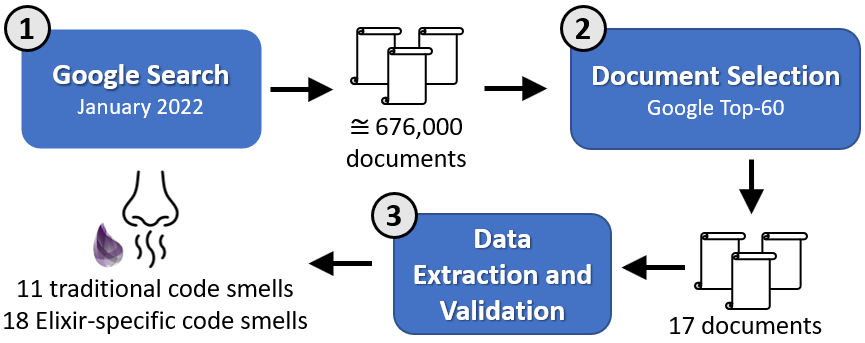}
  \caption{Overview of the grey literature methods}
  \label{fig:greyMethods}
  \Description{}
\end{figure}

Figure~\ref{fig:greyMethods} summarizes the steps we followed in the grey literature review. We also detail these steps in the following paragraphs.

\noindent \textit{1) Google Search:} 
According to Garousi et al.~\cite{GAROUSI2019}, when defining the keywords for a Google search it is important to perform preliminary searches to calibrate the queries, in order to combine synonyms or to exclude specific terms that might affect the results. After this initial calibration, we define the following search query:\\[-0.2cm]

\begin{lstlisting}[language=elixir]
("Elixir") AND 
("code smell" OR "code smells" OR "bad smell" OR "bad smells" OR "anti-pattern" OR "anti-patterns" OR       "antipattern" OR "antipatterns" OR "bad-practice" OR  "bad-practices" OR "bad practice" OR "bad practices")
\end{lstlisting}

This query includes possible code smells synonyms, both in singular and in plural. This was done to mitigate the risk that some desired discussions would be ignored. The final search was carried out in January 2022.\\[-0.2cm]


\noindent \textit{2) Document Selection:} As approximately 676,000 documents were retrieved, it would be impractical to analyze all of them. This is a recurring problem in grey literature reviews, so it is necessary to limit the number of documents to be analyzed \cite{GAROUSI2019}. Hence, we decided to select the top-60 documents returned by Google, i.e.,~the first six pages of results. These documents were read by the two authors in order to select only those that are indeed related to our research questions. After this step, 17 documents were selected, which we refer to as D1 to D17. The complete list of documents is available at: \url{https://doi.org/10.5281/zenodo.6025010}\\[-0.2cm]

\noindent \textit{3) Data Extraction and Validation:} The documents selected in the previous step were analyzed in detail by the first author, in order to retrieve sentences discussing traditional or novel code smells.
The extracted sentences were then validated by the second author. Only 2 out of 17 documents---D4 and D15---did not have code smells extracted due to a lack of agreement between the authors.  

\section{Results}
\label{sec:results}

\subsection{Do Elixir developers discuss traditional code smells?}

In the grey literature, we found discussions about 11 traditional code smells, as shown in Table~\ref{tab:traditionalSmellsGreyLit}. 

{\small
\begin{table}[!h]
  \caption{Traditional code smells (grey literature)}
  \label{tab:traditionalSmellsGreyLit}
  \begin{tabular}{p{4.0cm}p{2.5cm}p{0.5cm}}
    \toprule
    \textbf{Code Smell in Elixir} & \textbf{Documents}& \textbf{\#}\\
    \midrule 
    Comments & D1, D10, D12, D14 & 4\\
    Long Parameter List & D1, D16 & 2\\
    Feature Envy & D1, D6 & 2\\
    Shotgun Surgery & D1, D17 & 2\\
    Duplicated Code & D1 & 1\\
    Long Function & D1 & 1\\
    Large Class & D1 & 1\\
    Inappropriate Intimacy & D1 & 1\\
    Divergent Change & D1 & 1\\
    Speculative Generalization & D1 & 1\\
    Primitive Obsession & D3 & 1\\
  \bottomrule
\end{tabular}
\end{table}
}

The most discussed traditional smell is \textit{Comments} (4 documents). In D1, for example, the author argues that using comments to document code in Elixir is a bad practice:\\[-0.2cm] 

\noindent \textit{"Some people argue that any comment is a code smell, but this is not true. This is about comments that explain what the code does. [Instead] in Elixir we have a strong relationship with code documentation by using doctest. The documentation will describe how to use [a function], bring examples of use [...]. We can access all this via terminal."} \\[-0.2cm]

For D1's author, if the programmer's objective is to explain the purpose of a function (including the purpose of its parameters, return value, test samples, and so on), this should be done via {\tt doctest} in Elixir. When {\tt doctest} is used for code documentation, it is possible to access all the documentation quickly using Elixir's interactive shell (IEx) helper.\footnote{\href{https://hexdocs.pm/iex/1.13/IEx.Helpers.html}{https://hexdocs.pm/iex/1.13/IEx.Helpers.html}} Also, the usage examples provided for a function can be automatically executed as unit tests, something that is not possible using source code comments.


Although D17's author did not explicitly use the name \textit{Shotgun Surgery} in a document about implementing microservices in Elixir, he clearly refers to situations where particular code modifications require many small changes in different files: 

\noindent \textit{“
[When using microservices we]
need to be able to deploy independently. [Despite that] tight coupling could be found through shared libraries forcing an upgrade throughout the system. Or [microservices] could be coupled through a database schema where many services need to upgrade after a schema change.”} \\[-0.2cm]

We also found discussions about other smells, such as  \textit{Long Parameters List} (in D16):\\[-0.2cm]

\noindent \textit{“A long parameters list is one of many potential bad smells [...]. In object-oriented languages like Ruby or Java, we could easily define classes that help us solve this problem. Elixir does not have classes but because it is easy to extend, we can define our own types.”}\\[-0.2cm]

By analyzing  Table~\ref{tab:traditionalSmellsGreyLit}, we can also conclude that eight (out of 17) documents have discussions on traditional code smells. However, a single document (D1) concentrates most discussions. This document discusses all, except one, smells detected in our review. Moreover, six (out of 11) smells are only discussed in this specific document.

\subsection{Do Elixir developers discuss other smells?}

In the grey literature, we found discussions about 18 Elixir-specific code smells, which are
summarized in Table~\ref{tab:specificSmellsGreyLit3}. The most discussed one is \textit{Exceptions for control-flow} (2 documents). All other Elixir-specific smells are discussed only once each. 

To facilitate their discussion, we classify these smells into two different groups: \textit{design-related smells} (10 smells) and \textit{low-level concerns smells} (8 smells). Basically, we classify as \textit{design-related} smells that are more complex, that affect a coarse-grained  code element, and that are more difficult to detect. On the other hand, \textit{low-level concerns smells} are more simple and affect a small part of the code.

For example, according to D10's author, although multi-clause functions in Elixir is a powerful resource to keep the code interface small and organized, the abuse of this resource can make the code difficult to understand, characterizing the \textit{Complex multi-clause function} design-related smell, described as follows:  \\[-0.2cm]

\noindent \textit{“In Elixir, we can use multi-clause functions to group functions together using the same name. [However] when we start adding and mixing more pattern matchings and guard clauses [...] trying to squeeze too many business logics into the function definitions, the code will quickly become unreadable and even harder to reason with.[...]”}\\[-0.2cm]

As another example, the low-level concern smell called \textit{Working with invalid data} is described in D5 as follows:\\[-0.2cm]

\noindent \textit{“Elixir programs should prefer to validate data as close to the end-user [...] so the errors are easy to locate and fix. [When this is not done] if the user supplies an invalid input, the error will be raised deep inside [the function], which makes it confusing for users. [...] when you don't validate the values at the boundary, the internals [of the function] are never quite sure which kind of values they are working with.[...]”}\\[-0.2cm]

By analyzing  Table~\ref{tab:specificSmellsGreyLit3}, we can draw that eight (out of 17) documents discuss Elixir-specific smells. Only one document (D10) discusses both traditional smells and Elixir-specific smells simultaneously. Furthermore, D5 concentrates the highest number of discussions about Elixir-specific smells, since half of these new smells are detected in this document.

{\small
\begin{table*}[!t]
\renewcommand{\arraystretch}{1.13}
  \caption{Elixir-specific code smells (grey literature)}
  \label{tab:specificSmellsGreyLit3}
  \begin{tabular}{p{0.2cm}p{3.85cm}p{11.54cm}p{0.9cm}}
    \toprule
    & \multicolumn{1}{p{3.85cm}}{\textbf{Smell}} & \multicolumn{1}{p{11.54cm}}{\textbf{Description}} &
    \multicolumn{1}{p{0.9cm}}{\textbf{Docs}}\\
    \midrule

    \parbox[t]{2mm}{\multirow{10}{*}{\rotatebox[origin=c]{90}{Design-related smells}}} & 
    
    GenServer Envy & Using a Task or Agent but handling them like GenServers & D8 \\
    
    & Agent Obsession & When the responsibility for interacting with an Agent process is spread across the system & D8\\
    
    & Unsupervised process & Library that creates process outside a supervision tree, not allowing users to fully control their apps & D5 \\
    
    & Large messages between processes & Processes that exchange long messages frequently & D13 \\
    
    & Complex multi-clause function & Function with many guard clauses and pattern matching & D10 \\
    
    & Complex API error handling & Function that handles a large number of error types returned by an API endpoint, making it complex & D2 \\
    
    & Exceptions for control-flow & A library that forces clients to handle control-flow exceptions & D5, D11 \\
    
    & Untested polymorphic behavior & Function with a generic Protocol type parameter, but that does not have guards verifying its behavior & D7 \\
    
    & Code organization by process & Library unnecessarily organized as a process, instead as modules and functions & D5 \\
    
    & Data manipulation by migration & Low cohesive module that performs both data and structural changes in a DB schema via Ecto.Migration & D9 \\
    
    \midrule 
    \parbox[t]{2mm}{\multirow{9}{*}{\rotatebox[origin=c]{90}{Low-level concerns smells}}} & 
    
    Working with invalid data & Function that does not validate its parameters and therefore can produce non-predicted behaviors & D5 \\
    
    & Map/struct dynamic access & Function that dynamically accesses values from a non-existent field in a struct or map & D7 \\
    
    & Unplanned value extraction & Function that does not force a crash when an incorrect value is extracted from a URL query string & D7 \\
    
    & Modules with identical names & Modules with conflicting names, preventing their simultaneous load & D5 \\
    
    & Unnecessary macro & Using macros instead of functions and structs & D5 \\
    
    & App configuration for code libs & Function that uses global configuration mechanisms, preventing clients from reusing it flexibly & D5 \\
    
    & Compile-time app configuration & Function that uses module attributes to configure itself, therefore preventing run-time configurations & D5 \\
    
    & Dependency with "use" when an "import" is enough & Module that requires knowledge of its internal dependencies to understand its behavior & D5 \\
  \bottomrule
\end{tabular}
    {\centering *\textit{Note:} more details on Elixir-specific code smells are available at \url{https://github.com/lucasvegi/Elixir-Code-Smells} \par}
\end{table*}
}

\subsection{Does a well-known static code analysis tool for Elixir detect code smells?}

Code smell detection tools---including mainstream tools such as SonarQube\footnote{\href{https://www.sonarqube.org/features/quality-gate/}{https://www.sonarqube.org/}} and research-oriented tools such as DECOR~\cite{DECOR2010}---do not provide support to Elixir. Therefore, we decided to check whether the smells reported in  RQ1 and RQ2 are detected by Credo, which is currently the most popular static code analysis tool for Elixir. Credo performs the following checks: software design issues, code readability, refactoring opportunities, warnings, and consistency. 

We carefully analyzed Credo documentation (version 1.6.2), aiming to find warnings for the code smells unveiled in our research. 
Interestingly, most smells listed in RQ1 are not detected by Credo. In fact, only two traditional code smells---\textit{Duplicated code} and \textit{Long Parameter List}---are detected by this tool. Moreover, Credo also does not detect the more complex and design-related smells described in RQ2. 
Particularly, we concluded that only one Elixir-specific smell---\textit{Compile-time app configuration}---is  detected by Credo.
This finding represents an opportunity for extending tools such as Credo to detect code smells.

\section{Threats to Validity}
\label{sec:threats}

Since our methodology is based on a grey literature review, we analyzed documents that are not peer-reviewed. However, to reinforce the validity of our results, we carefully selected the documents returned by the search engine. Both paper authors did a preliminary reading of all the top-60 documents returned by Google, selecting for analysis only those that are relevant and are written by professionals who work with Elixir.

Another threat  concerns the format of the search query. In a grey literature review, there is a risk that important results will not be found due to missing keyword combinations in the search query. To mitigate this threat, as proposed by Garousi et al. \cite{GAROUSI2019}, we performed preliminary searches, adding and deleting some keywords.

The final threat relates to how we analyzed the smells detected by Credo. However, to minimize this threat, we have thoroughly reviewed all Credo documentation. Although this documentation is rich in details and examples, we might have missed some smells.




\section{Related Work}
\label{sec:relatedWork}

To the best of our knowledge, this is the first work that catalog code smells to a modern functional programming language, more specifically for Elixir. As code smells are context-sensitive, other studies have cataloged code smells for specific contexts, such as Android~\cite{Hecht2015}, iOS~\cite{iosSmells}, JavaScript~\cite{Saboury2017}\cite{JSNOSE}, CSS~\cite{cssSmells}, and Puppet~\cite{PuppetSmells}.

As in our work, these papers also used grey literature as a source to discover context-specific code smells. In contrast, they also search for smells in real projects, which we plan to conduct in the near future, as described next.





\section{Future Work}
\label{sec:futureWork}

As future work, we plan to extend and validate our list of smells by mining and analyzing issues, pull requests and commits of Elixir open-source projects. We also plan to conduct surveys and interviews with Elixir developers. Next, we plan to invest on tools for detecting smells in Elixir, possibly by extending current static analysis tools, such as Credo. Finally, we have plans to investigate code smells in other modern functional programming languages, such as Clojure.




\balance

\bibliographystyle{ACM-Reference-Format}
\bibliography{sample-base}

\end{document}